\documentclass[reprint,twocolumn, amsmath,amssymb, aps]{revtex4}

\usepackage{graphicx}
\usepackage{dcolumn}
\usepackage{bm}
\usepackage{babel}
\usepackage{color}
\usepackage{soul}
\usepackage{ulem,xpatch}
\usepackage{empheq}
\usepackage{amsmath}
\usepackage{amssymb}

\begin{document}

\title{Modelling and observation of nonlinear damping in dissipation-diluted nanomechanical resonators} 

\author{Letizia Catalini}
\author{Massimiliano Rossi}
\altaffiliation{Current address: Photonics Laboratory, ETH Zürich, 8093 Zürich, Switzerland}
\author{Eric C. Langman}
\author{Albert Schliesser}
\email{albert.schliesser@nbi.ku.dk}

\affiliation{Niels Bohr Institute, University of Copenhagen, Blegdamsvej 17, 2100 Copenhagen, Denmark}
\affiliation{Center for Hybrid Quantum Networks (Hy-Q), Niels Bohr Institute, University of Copenhagen, 2100 Copenhagen, Denmark}
\date{\today}

\begin{abstract}
Dissipation dilution enables extremely low linear loss in stressed, high-aspect ratio nanomechanical resonators, such as strings or membranes. Here, we report on the observation and theoretical modelling of nonlinear dissipation in such structures. We introduce an analytical model based on von K\'arm\'an theory, which can be numerically evaluated using finite-element models for arbitrary geometries. We use this approach to predict nonlinear loss and (Duffing) frequency shift in ultra-coherent phononic membrane resonators. A set of systematic measurements with silicon nitride membranes shows good agreement with the model for low-order soft-clamped modes. Our analysis also reveals quantitative connections between these nonlinearities and dissipation dilution.
This is of interest for future device design, and can provide important insight when diagnosing the performance of dissipation dilution in an experimental setting. 
\end{abstract}

\maketitle
\section{Introduction}

In recent decades, micro- and nanomechanical systems have attracted widespread interest in science and technology \cite{Cleland_2003, Schmid_2016}.
They constitute outstanding sensors for force \cite{reinhardt_ultralow-noise_2016}
, mass  \cite{Chaste_2012}, radiation \cite{Yi_2013}, and temperature \cite{Chien_2018}, to name just a few examples.
Simultaneously, they are promising building blocks for future quantum technologies, such as microwave- or spin-to-optical quantum transducers \cite{Midolo_2018, Kurizki_2015} or quantum memories \cite{pechal_superconducting_2018}.

Low thermomechanical noise, and correspondingly long mechanical coherence times are crucial for these applications, and are typically limited by energy dissipation from the mode of interest.
Considerable effort has therefore gone into designing mechanical resonators with minimal dissipation, leading to great advances in recent decades.
Beyond mitigating all external losses, e.\ g.\ to the surrounding gas or the device  substrate, important progress was made in suppressing loss to internal degrees of freedom, such as two-level systems.  

By storing the majority of the mechanical mode's energy in a loss-less potential, dissipation dilution \cite{Gonz_lez_1994} has emerged as a successful strategy in this endeavour.
It can be utilized in highly-stressed nanomechanical string and membrane systems, whereby the elongation energy assumes the role of the loss-less potential \cite{Huang_1998, Verbridge_2007, Unterreithmeier_2010, Schmid_2011, Fedorov_2019}.
We have recently introduced an extension of this approach --- soft-clamping  --- to engineer mechanical resonance modes particularly conducive to dissipation dilution in phononic crystal membranes \cite{Tsaturyan_2017}. 

Soft-clamping has allowed realizing nanomechanical resonators with the highest $Q$-factors ($>10^8$) and $Qf$-products ($>10^{15}\,\mathrm{Hz}$) yet observed at room temperature \cite{Tsaturyan_2017, Ghadimi_2018, Reetz_2019, Catalini_2020}.
For the vast majority of mechanical systems, it is sufficient to consider dissipation in the linear regime, i.\ e.\ when the quality factor is independent of the displacement amplitude \cite{Cleland_2003}.
Some instances of nonlinear dissipation in nanomechanical systems have been reported, for example in nano-resonators made from diamond \cite{Imboden_2013}, carbon nanotubes and graphene sheets \cite{Eichler_2011}, but without providing a clear explanation as to the origin of this effect.
Here, we investigate nonlinear effects in soft-clamped membrane resonators with very high Q-factors.
Whereas the Duffing frequency shift has been observed in stressed nanomechanical resonators before \cite{Fong_2012, Hocke_2014,Defoort_2012},
we focus on nonlinear damping here \cite{Zaitsev_2011, Catalini_2020}.

Starting from a full 3D model (von K\'{a}rm\'{a}n theory), we derive analytical expressions for both the Duffing frequency shift and nonlinear damping, similar to what has been derived for a string [2].
This analysis furthermore reveals strong connections to dissipation dilution, both being linked to geometric nonlinearities.
A series of systematic experiments yields good quantitative agreement with the model for low-order soft-clamped modes.
We thereby establish not only a means to quantitatively predict nonlinear losses --- as relevant e.\ g.\ for parametric sensing protocols \cite{Ko_ata_2020} --- but also introduce a new experimental tool for assessing dissipation dilution.

\section{Model}
We describe the motion of a thin membrane, of thickness $h$. The xy-plane coincides with the one of the undeformed membrane.

The displacement of the mass element located at position $\mathbf{r}(x, y, z)$ is quantified by the vector $u_i$, where Latin indexes represent the three directions $x, y, z$. The membrane's deformation due to the motion of the mass elements is expressed by the strain tensor $\varepsilon_{ij}=\left(\partial_j u_i+\partial_i u_j+\partial_i u_z\partial_j u_z\right)/2$.
The deformation induces stresses within the structure, described by the stress tensor $\sigma_{ij}$. We consider  elastic materials, for which the induced stresses are linear in the strain tensor and Hooke's law holds \cite{Landau1970}.
For thin membranes with no external loads, stress components associated with the $z$-direction are negligible, i.e. $\sigma_{iz}=0$.
The membrane's displacement can then be decomposed into the out-of-plane displacement, $u_z(x,y,z)\equiv w(x,y)$, and the in-plane displacement $u_\alpha(x,y,z)=v_\alpha(x,y)-z\partial_\alpha w(x,y)$, where Greek indexes represent the in-plane coordinates $x$ and $y$, and $v_\alpha$ is the in-plane displacement. %
For amplitudes relevant to this work, the in-plane displacement components are negligible with respect to the out-of-plane components (confirmed with FEM simulations). Therefore, we apply the so-called out-of-plane approximation and neglect them \cite{Atalaya_2008}. Furthermore, our model includes a static in-plane deformation, $\varepsilon_0 (x,y)$, representing the static strain required for dissipation dilution. The shear components of the stress tensor, $(\sigma_0)_{xy}$, are negligible in the structures considered. Within these approximations, we can write the strain and stress tensors, according to von K\'{a}rm\'{a}n theory \cite{Landau1970,Atalaya_2008}, as
\begin{align}
    \varepsilon_{\alpha\beta}&=\varepsilon_0\delta_{\alpha\beta}-z\partial_{\alpha\beta}w+\frac{1}{2}\partial_{\alpha}w\partial_{\beta}w,\\
    \sigma_{\alpha\beta}&=\frac{E}{1-\nu^2}\left[(1-\nu)\varepsilon_{\alpha\beta}+\nu\varepsilon_{\gamma\gamma}\delta_{\alpha\beta}\right],
\end{align}
where $E$ is the Young's modulus, $\nu$ the Poisson's ratio, $\delta_{\alpha\beta}$ the Kronecker delta and the repeated Greek indices are summed over.

To introduce dissipation, we assume a time delay $\tau$ in the stress-strain relation, as a phenomenological model for microscopic relaxation processes intrinsic to the resonator material \cite{Schmid_2016}. For small and constant time delay (relative to mechanical period), the stress tensor can be approximated as $\sigma(t)=H[\varepsilon(t+\tau)]\approx H[\varepsilon(t)]+\tau H[\dot{\varepsilon}(t)]$, where $H$ is a linear functional expressing Hooke's law. 
The dissipation arises from the additional term proportional to $\tau$. The equation of motion for out-of-plane displacement is
\begin{subequations}
    \label{eq:eom_vk}
    \begin{align}
        \rho h \ddot{w}-\partial_{\alpha\beta}M_{\alpha\beta}-\partial_{\beta}\left(N_{\alpha\beta}\partial_{\alpha}w\right)=0,\label{eq:eom_1}\\
        \partial_\beta N_{\alpha\beta}=0,
    \end{align}
\end{subequations}
where the stress resultants $N_{\alpha\beta}$ and $M_{\alpha\beta}$ are the shear force components and bending momenta, respectively. They are given by $N_{\alpha\beta}=\int \sigma_{\alpha\beta}dz$ and $M_{\alpha\beta}=\int z\sigma_{\alpha\beta}dz$, where each integral is performed over the membrane thickness. As both stress resultants contain terms proportional to the time delay $\tau$, they generate both linear and nonlinear dissipative processes, as we shall see.
Typically, equations~\eqref{eq:eom_vk} are difficult to solve due to nonlinearity in the displacement $w$. For small $w$, we neglect nonlinear terms and retrieve a solvable linear equation, yielding a set of normal modes $w_\eta(x, y, t)=\phi_\eta(x, y)u_\mu(t)\delta_{\eta\mu}$, where $\eta$ indexes different vibrational modes and we have separated the time-dependent mode amplitude $u_\mu(t)$ from the dimensionless transverse spatial profile $\phi_\eta(x, y)$.
We use this set of transverse modes as a basis to expand a solution of the full nonlinear equation of motion, that is $w(x, y, t)=\phi_\eta(x, y)u_\eta(t)$. 
We insert this ansatz in Eq.~\eqref{eq:eom_1}, then project it into a single transverse mode, $\phi_i$, by applying the Galerkin method together with a single mode approximation that neglects intermodal coupling \cite{Younis_2011}.
Finally, we obtain the following effective nonlinear equation for the temporal mode $u_i$
\begin{equation}\label{eq:eff_eom_singlemode}
    \ddot{u}_i+\Gamma_i \dot{u}_i+\gamma_i^\mathrm{nl}u_i^2\dot{u}_i+\Omega_i^2u_i+\beta_i u_i^3=0,
\end{equation}
which describes a damped Duffing resonator \cite{Hocke_2014, Fong_2012}, including a nonlinear damping term \cite{Catalini_2020, Zaitsev_2011, Polunin_2016, Antoni_2012,Gusso_2020}.
The effective parameters in Eq.~\eqref{eq:eff_eom_singlemode} are defined as
\begin{subequations}
\label{eq:main}
\begin{align}
    \Omega_i^2&=m_\text{eff}^{-1}\int\phi_i\left[D\partial_{\alpha\alpha\beta\beta}\phi_i-h\sigma_0\partial_{\alpha\alpha}\phi_i\right]dA,\label{eq:omega}\\
    \beta_i&=\frac{k_1}{2}\int\phi_i\left(\partial_{\alpha\beta}\phi_i\partial_{\alpha}\phi_i\partial_{\beta}\phi_i+k_2\partial_{\alpha\alpha}\phi_i\partial_{\beta}\phi_i\partial_{\beta}\phi_i\right)dA\label{eq:beta},\\
    \Gamma_i&=\tau D\,m_\mathrm{eff}^{-1} \int\phi_i\partial_{\alpha\alpha\beta\beta}\phi_i dA,\label{eq:gamma}\\
    \gamma_i^\mathrm{nl}&=k_1\tau\int\phi_i\left(\partial_{\alpha\beta}\phi_i\partial_{\alpha}\phi_i\partial_{\beta}\phi_i+k_2\partial_{\alpha\alpha}\phi_i\partial_{\beta}\phi_i\partial_{\beta}\phi_i\right)dA\label{eq:gammanl},
\end{align}
\end{subequations}
where the integrals extend over the whole membrane surface,
$m_\text{eff}=\rho h \int \phi_i^2 dA$ is the effective mass,
$D=Eh^3/(12(1-\nu^2))$ the flexural rigidity and we have introduced the constants $k_1=-hE/(m_\mathrm{eff}(1-\nu^2))$ and $k_2=\nu/(1-\nu)$. Notably, the two nonlinear terms are purely geometric effects and they are not introduced by the material itself.

As expected, we find both the linear ($\Gamma_i$) and nonlinear ($\gamma_i^{\mathrm{nl}}$) dissipation proportional to the lag time $\tau$.
In the context of dissipation dilution, the linear dissipation is commonly expressed as 
\begin{equation}
    \Gamma_i=\frac{1}{D_{Q,i}}\frac{\Omega_i}{Q_\mathrm{intr}},
\end{equation}
where $D_{Q,i}\gg 1$ is the dissipation dilution factor, determined by the geometry of mode $i$ \cite{Gonz_lez_1994, Huang_1998, Verbridge_2007, Unterreithmeier_2010, Schmid_2011, Tsaturyan_2017, Fedorov_2019}. 
The resonator's material properties enter via the intrinsic quality factor
$Q_\mathrm{intr}=(\Omega_i \tau)^{-1}$, which we use interchangeably with the material's loss angle $\theta_\mathrm{lin}=Q_\mathrm{intr}^{-1}$.  
In dissipation-diluted devices, one expects to measure enhanced (linear) quality factors
\begin{equation}\label{eq:diluted_q}
    Q_\mathrm{meas}:=\frac{\Omega_i}{\Gamma_i} = D_{Q,i} Q_\text{intr}=D_{Q,i} \theta_\text{lin}^{-1},
\end{equation}
compared to resonators made from the same material in absence of dissipation dilution (e.\ g.\  unstressed). 
In this setting,  $D_{Q,i}$ and $Q_\text{intr}$ are not separately accessible through measurement.

Turning to nonlinear effects, we note the Duffing frequency shift ($\beta_i$) and nonlinear damping ($\gamma_i^\mathrm{nl}$) depend on the mode pattern identically.
Yet their ratio depends on the lag time $\tau$ and thence on the intrinsic loss.
We therefore introduce the nonlinear loss angle 
\begin{equation}\label{eq:def_theta_nl}
    \theta_\mathrm{nl}:=\frac{\gamma_i^{\mathrm{nl}}\Omega_i}{2\beta_i},
\end{equation}
which notably depends only on quantities which can be experimentally measured through large-amplitude excitation (see below).
Importantly, eqs.~\eqref{eq:main} suggest 
\begin{equation}\label{eq:nl_eq_lin_hyp}
    \theta_\mathrm{nl}=\Omega_i \tau=\theta_\mathrm{lin},
\end{equation}
which would imply access to the intrinsic linear damping through measurement of a device's nonlinear properties.

\section{Experimental results}

\begin{figure}[htb]
\center
\includegraphics[scale=1]{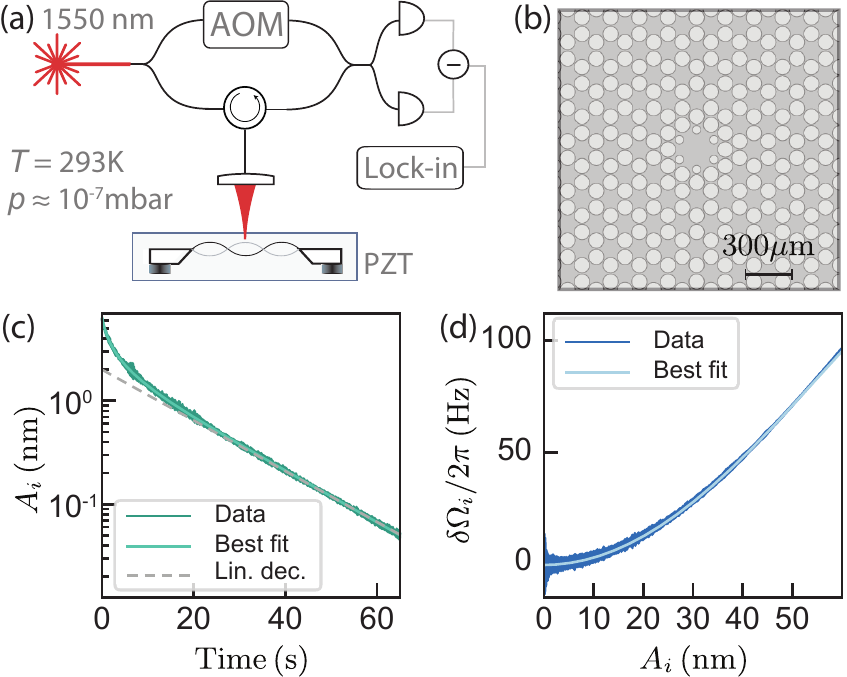}
\caption{(a) Optical interferometer for displacement measurement (AOM: acousto-optic modulator, PZT: piezoelectric actuator). The detection is realized with a heterodyne receiver. 
(b) Soft-clamped membrane pattern.  
(c) Nonlinear amplitude decay for 19-nm-thick membrane and corresponding fit. Linear exponential decay (gray) is extrapolated to highlight deviation arising from nonlinear damping. (d) Duffing frequency shift as a function of displacement amplitude and corresponding fit. The abscissa is the fit result from (c).}
\label{f:fig2}
\end{figure}

Our experimental subjects are highly-stressed 3.6mm~$\times$~3.6mm soft-clamped $\text{Si}_3\text{N}_4$ membrane resonators \cite{Tsaturyan_2017}, shown in Fig.~\ref{f:fig2}b, operated at room temperature and pressures lower than $10^{-7}$~mbar to reduce gas damping to a negligible value.
We expect intrinsic material dissipation to dominate and the theoretical framework derived above to apply since radiation losses of the vibrational defect modes are shielded by the honeycomb phononic crystal pattern. 
Nonlinear phenomena are present during free decay evolution as amplitude-dependent damping and shift of the mechanical resonance frequency.
To observe these effects, we employ ringdown techniques and measure mechanical displacement with a fiber-based optical Mach-Zender interferometer (Fig.~\ref{f:fig2}a).
Then, we stop driving and monitor the displacement decaying with a heterodyne detector, from which we extract both the displacement amplitude $A_i$ and phase, thus the instantaneous frequency $\Omega_i'$.
The displacement amplitude and frequency evolve according to \cite{Catalini_2020}:
\begin{align}
    \delta \Omega_i(t)&\equiv \Omega_i-\Omega_i'=\frac{3}{4}\omega_i^{sD}A_i^2(t), \label{eq:backbone}\\
    A_i(t)&=\frac{A_{i_0}e^{-\frac{\Gamma_i}{2}t}}{\sqrt{1+\frac{\gamma_i^\mathrm{nl}}{4\Gamma_i}A_{i_0}^2\left(1-e^{-\Gamma_i t}\right)}}. \label{eq:nl-decay}
\end{align}
Equation~\eqref{eq:backbone} is the standard backbone equation \cite{Fong_2012,Hocke_2014}, while eq.~\eqref{eq:nl-decay} describes non-exponential decay, induced by the nonlinear damping \cite{Polunin_2016,Zaitsev_2011}.
We have also introduced the Duffing shift per displacement, $\omega_i^\text{sD}:=\beta_i/(2\Omega_i)$, such that the nonlinear loss angle defined in eq.~\eqref{eq:def_theta_nl} becomes $\theta_\text{nl}^{-1} = 4 \omega_i^\text{sD}/\gamma_i^\text{nl}$

In Fig.~\ref{f:fig2}c and d, we show an example of nonlinear amplitude decay and frequency shift as a function of displacement amplitude. The nonlinear parameters are extracted from the best fit.
Importantly, the values of both $\omega_i^\text{sD}$ and $\gamma_i^\text{nl}$  depend on the displacement calibration \cite{Catalini_2020}, however their ratio $\theta_\text{nl}$ is independent of that.
Notice that deviation from linear decay starts appearing for displacement amplitudes comparable to the membrane thickness \cite{Gusso_2020}, as expected when the bending at the membrane edges has been eliminated (soft-clamping).
From the amplitude decay fit, we extract the linear damping rate $\Gamma_i$ and  calculate $Q_\text{meas}$.

\begin{figure*}[htb]
\center
\includegraphics[scale=1]{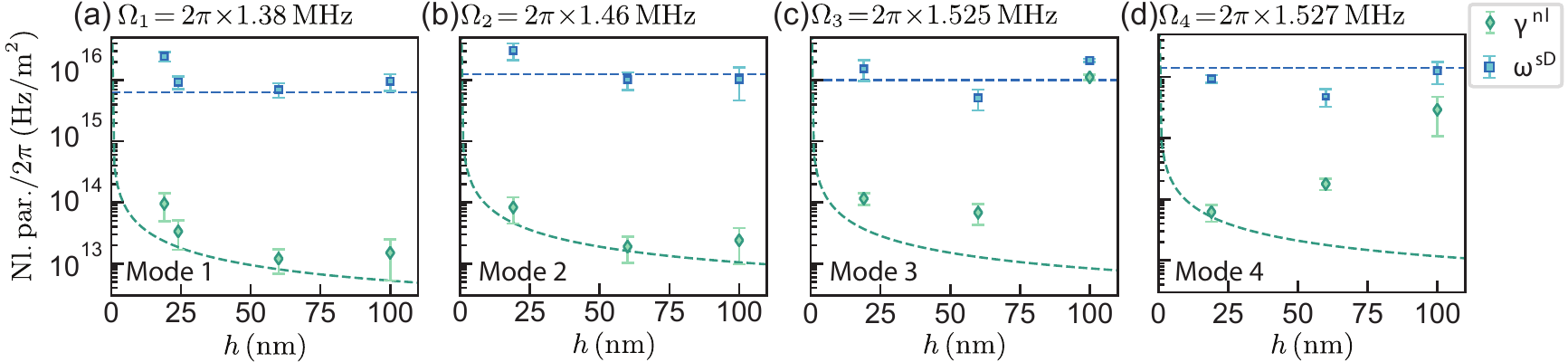}
\caption{Nonlinear parameters. (a)-(d) Measured nonlinear parameters as a function of membrane thickness $h$. Blue (green) points are the Duffing (nonlinear damping) parameters, $\omega_i^{sD}$ ($\gamma_i^{nl}$), estimated from the median of each statistical ensemble. Dashed lines represent simulated values for the Duffing (blue) and nonlinear damping (green) parameters.}
\label{f:fig3}
\end{figure*}

We perform 5 ringdown measurements on every individual mode, then average the results.  
Ringdowns are discarded if relative errors on fit parameters are greater than $10$~\%, using a 95~\% confidence interval.
For each membrane, we characterize the nonlinear parameters of four defect modes lying in the bandgap from $1.30$~MHz to $1.55$~MHz (see supplementary). Statistics are collected from 6 to 12  nominally identical membranes. 
Due to the presence of outliers within this ensemble, we evaluate and report the median and median absolute deviation as robust estimators of the ensemble statistics \cite{Pham_Gia_2001}. 
The measurement protocol stated above is repeated for membranes of different thickness, with all nonlinear parameters reported in Fig.~\ref{f:fig3}.
(The values obtained for mode 3 of two membranes were discarded since a negative Duffing shift was observed.) 

We simulate the parameters according to eqs.~\eqref{eq:beta} and \eqref{eq:gammanl}, where the transverse profile of each mode is obtained by FEM simulations.
For modes 1 and 2 we find very good agreement for all investigated thicknesses in the range 19-100~nm. 
For thicker membranes, we observe excess nonlinear damping in modes 3 and 4, though the Duffing nonlinearity still matches predictions. 
Possible origins of this deviation are discussed below.

\begin{figure*}[htb]
\center
\includegraphics[scale=1]{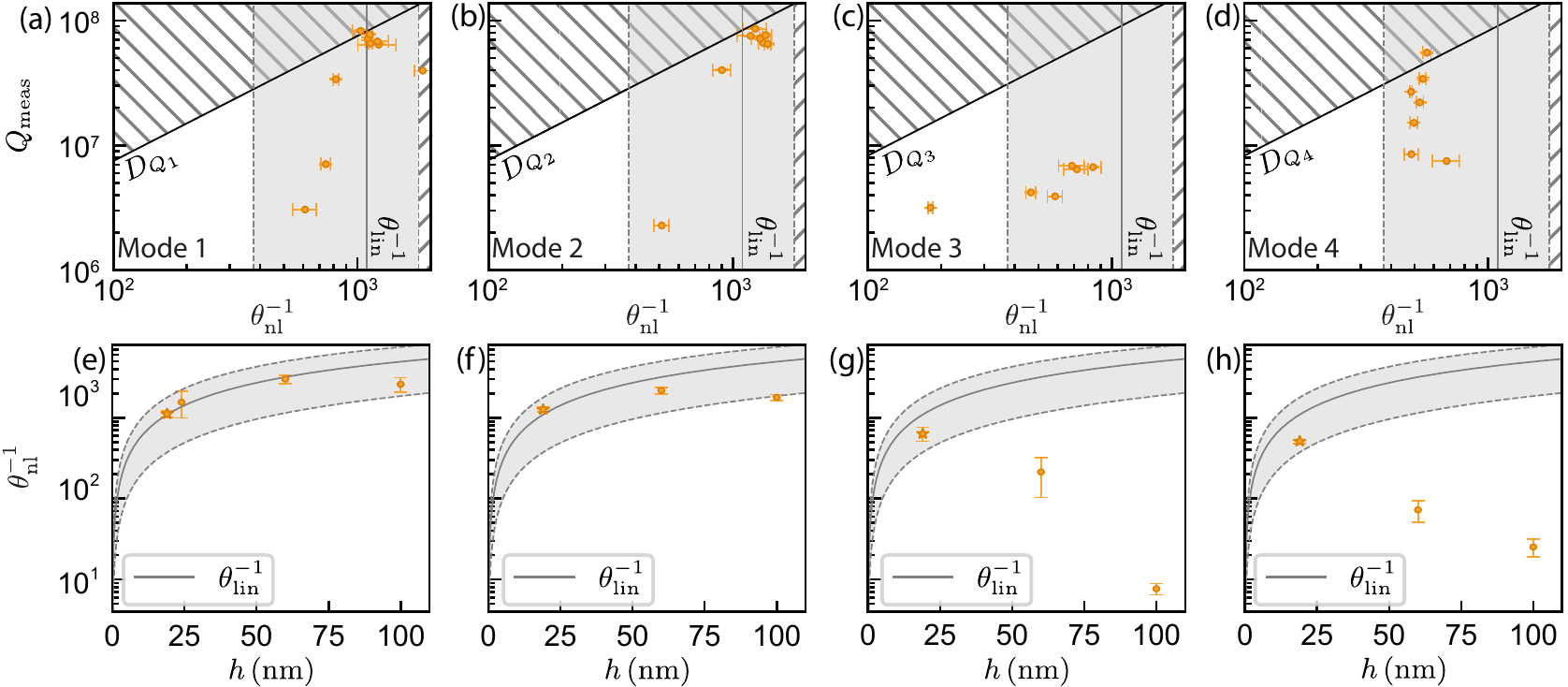}
\caption{Nonlinear loss angles. (a)-(d) Measured quality factors against $\theta_\text{nl}^{-1}$. The gray line is the expected $\theta_\text{lin}^{-1}$ \cite{Villanueva_2014} and gray area the uncertainty in that value. The black line is the limit $Q_\text{meas}\leq D_Q \theta_\text{lin}^{-1}$ with a dissipation dilution factor $D_Q$ obtained from simulations. The hatched area is inaccessible under the most obvious sources of excess dissipation. (e)-(h) Measured $\theta_\text{nl}^{-1}$ as a function of the membrane thickness.
The stars represent the median of the points show in the panels above. The gray line is the $\theta_\text{lin}^{-1}$ as a function of thickness, expressed in eq.~\eqref{eq:qintr}, whereas the gray area reflects the uncertainty in the parameters.}
\label{f:fig4}
\end{figure*}

By comparing nonlinear and linear losses experimentally, we now examine the hypothesis of eq.~\eqref{eq:nl_eq_lin_hyp}. 
In Fig.~\ref{f:fig4}, the measured linear quality factors $Q_\mathrm{meas}$ are plotted against extracted $\theta_\text{nl}^{-1}$.
Per our hypothesis, the measured quality factor cannot exceed the function $ Q_\mathrm{meas}\stackrel{!}{=}D_Q \theta_\mathrm{nl}^{-1}$, which follows from eq.~\eqref{eq:diluted_q} if eq.~\eqref{eq:nl_eq_lin_hyp} is correct. The dilution factor for each mode is obtained from FEM simulations, as described earlier \cite{Tsaturyan_2017}. This relation should hold independent of material quality, which may vary between fabrication runs. 

Linear loss angles of silicon nitride thin films have been extracted from a large number of resonator data reported in the literature \cite{Villanueva_2014}.
The expected range is represented by the gray area, where the gray line is the average value.
If the hypothesis eq.~\eqref{eq:nl_eq_lin_hyp} holds, $\theta_\text{nl}$ (abscissa in Fig.~\ref{f:fig4} ) should also fall in this range.

Although each point should ideally fall between the two lines, additional losses can be introduced during fabrication and handling. 
The data points would then have larger loss angle but still lie on the oblique line.
On the other hand,  imperfect dissipation dilution (e.~g.~ damaged structures,  residual gas damping, or radiation loss) would lead to data points lying below the oblique line. 
The regions not explained by these mechanisms are shown as hatched. We observe that the vast majority of the measurements are within the expected region. For the first two modes, most points lie close to the intersection, corroborating our hypothesis. The points' location with respect to the intersection then gives an indication of possible imperfections in the sample.

Nonlinear loss angle measurements for different thicknesses are shown in Fig.~\ref{f:fig4}e--h. 
We compare this to the phenomenological model of the linear loss angle as a function of membrane thickness:
\begin{equation}\label{eq:qintr}
    \theta_\text{lin}(h)=Q_\mathrm{intr}^{-1}(h)=Q_\mathrm{vol}^{-1}+(\beta_s h)^{-1},
\end{equation}
where $Q_\mathrm{vol}=(2.8\pm 0.2)\times10^4$ and $\beta_s=(60\pm 40)\,\mathrm{nm}^{-1}$ \cite{Villanueva_2014}. 
Measured $\theta_\text{nl}$ for modes 1 and 2 agree with the model for the $\theta_\text{lin}$, within error, supporting our hypothesis. However, the data for mode 3 and 4 show an evident deviation for larger thickness. This is consistent with the excess nonlinear damping observed in Fig.~\ref{f:fig3}c and d. The source of this excess nonlinear damping is unclear.
Since the isolation provided by the phononic shield for these two modes is in general worse  \cite{Tsaturyan_2017}, we speculate that this can lead to nonlinear energy exchange mediated by vibrational modes of the supporting silicon frame \cite{patil_thermomechanical_2015}.

\begin{figure}[ht!]
\center
\includegraphics[scale=.9]{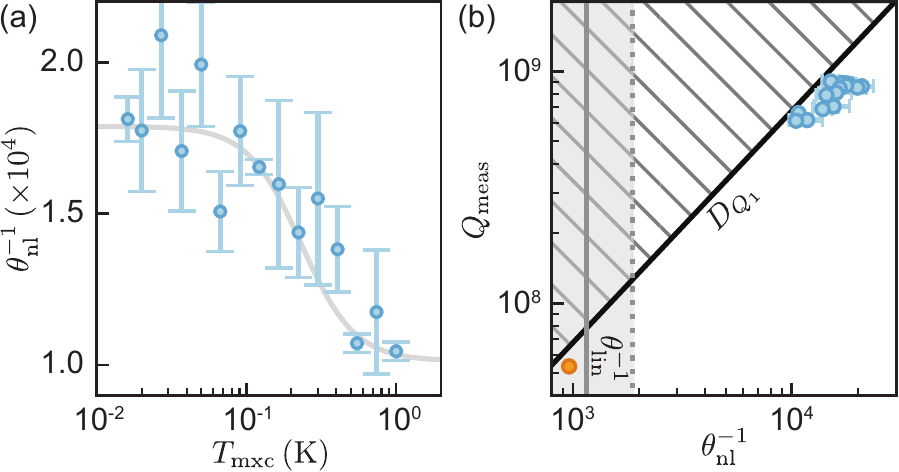}
\caption{Nonlinear loss angles at different temperatures. (a) Measured $\theta_\text{nl}$ as a function of the cryostat temperature, $T_\text{mxc}$. The gray line is a polynomial fit, roughly showing the behavior.
(b) Measured quality factors versus $\theta_\text{nl}$, taken at room temperature (orange) and cryogenic temperatures (blue). The gray line is the expectation value of $\theta_\text{lin}^{-1}$ at room temperature, and the gray area reflects uncertainty \cite{Villanueva_2014}. The black line is the simulated quality factor, from the dissipation dilution factor $D_{Q_1}$. Error bars are the mean absolute deviation among 3 repetitions.}
\label{f:fig5}
\end{figure}

Lastly, we characterize the intrinsic losses as a function of the temperature (cf.\ Fig.\ \ref{f:fig5}). 
Ringdown measurements were performed on mode 1 of a 19-nm-thick membrane inside a dilution refrigerator, at temperatures ranging from $20$~mK to $1$~K.
We use $100$~nW of optical power at $830$~nm impinging on the membrane, to minimize heating from absorption of optical radiation \cite{Page_2020}.
An additional reference measurement is taken at room temperature.
As reported previously \cite{yuan_silicon_2015, faust_signatures_2014,fischer_optical_2016,Page_2020},  we observe an increase of linear quality factor with decreasing temperature.
We find $\theta_\text{nl}$ decreases in unison with the linear loss over nearly 4-orders of magnitude span in temperature, in line with our hypothesis. 
Our new analysis also gives two insights of potential use for further experimental optimization: (a) saturation of the decrease in $\theta_\text{nl}$ at around 100~mK suggests that the sample might not thermalize properly to lower temperatures,  (b) near $10^9$ the measured linear $Q$-factors stop following $\theta_\text{nl}$, suggesting the presence of additional linear, undiluted dissipation. 

\section{Conclusion}

Our work shed lights on the origin of nonlinear damping in dissipation-diluted nanomechanical resonators.
We have developed an analytic theory based on a continuum elastic model for large deflections of a thin membrane. 
The geometric nonlinearity arising from the material elongation modifies both the conservative and dissipative dynamics, in the form of Duffing frequency shifts and a nonlinear damping.
We observe these nonlinear effects in soft-clamped, ultracoherent membrane resonators and find good agreement with our model.

We introduce the nonlinear loss angle $\theta_\text{nl}$ and show that it can be extracted from ringdown measurements without displacement calibration. 
Our model hypothesizes $\theta_\text{nl}$ is equal to the linear loss angle, which is otherwise not separately accessible by measurement.
We find substantial evidence supporting this hypothesis across a wide array of mode shapes, geometric parameters, and temperatures.
These insights deepen our understanding of nonlinear behaviour in this important class of nanomechanical resonators, and can guide design of future generations of ultracoherent mechanical sensors \cite{Tsaturyan_2017,Ghadimi_2018, Reetz_2019}, especially with regard to sensing protocols \cite{Ko_ata_2020}.
Finally, the tools developed here yield additional insight in the performance and loss contributions of dissipation-diluted resonators. 

\section*{Acknowledgments}
The authors acknowledge Y.~Tsaturyan for sample fabrication and D.~Mason for assistance with the interferometric nonlinear transduction. This work was supported by the Swiss National Science Foundation (grant nr. 177198), the European Research Council project Q-CEOM (grant nr. 638765), the Danish National Research Foundation (Center of Excellence “Hy-Q”), the EU H2020 FET proactive project HOT (grant nr. 732894), and the Novo Nordisk Foundation (grant nr. NNF20OC0061866).

\begin{widetext}
\appendix
\section*{Appendices}
\section{Extended Theory}\label{app:theory}
Here we present in detail the derivation of the equations of motion of a square membrane with thickness $h$ and side length $L$, when the out-of-plane displacement is comparable with the membrane thickness. The xy-plane coincide with the one of the undeformed membrane.\\
We start our derivation by describing the deformation of the membrane due to the oscillation. The displacement of the mass element located at the position $\mathbf{r}=(x,y,z)$ is quantified by the displacement vector $\mathbf{u}=(u_x,u_y,u_z)$. In the general case, the deformation of the membrane due to the motion of the mass elements is described by the generalized strain tensor $\varepsilon_{ij}=2^{-1}(\partial_ju_i+\partial_iu_j+\partial_iu_k\partial_ju_k)$, where $i$, $j$, $k$ stands for the three directions $x$, $y$, $z$ and the repeated indices are summed over. In the linear case the second order terms are neglected. Here we consider the case where the out-of-plane displacement vector component is as large as the membrane thickness, whereas the two in-plane displacement vector components are small. In this regime, the strain tensor considered is
\begin{eqnarray}
    \varepsilon_{ij}=\frac{1}{2}(\partial_ju_i+\partial_iu_j+\partial_iu_z\partial_ju_z).
\end{eqnarray}
From elasticity theory, we know that every time a material is deformed, a stress is generated within the structure, described by the stress tensor $\sigma_{ij}$. In our treatment we consider elastic material, i.e. material for which the induced stresses are linear function of the strain tensor and such a function is described by Hooke's law:
\begin{equation}
    \sigma_{ij}=\frac{E}{1+\nu}\left(\varepsilon_{ij}+\frac{\nu}{1-2\nu}\varepsilon_{kk}\delta_{ij}\right),
\end{equation}
where $E$ is the Young's modulus and $\nu$ the Poisson's ratio and $\delta_{ij}$ the Kronecker delta.\\
We consider a thin membrane, i.e. $L\gg h$, with no external loads and we assume that the stress components associated with the $z$ direction are negligible, $\sigma_{iz}=0$.
Within these conditions, the displacement vector is described by an out-of-plane displacement $u_z(x,y,z)=w(x,y)$ and the in-plane displacement $u_\alpha(x,y,z)=v_\alpha(x,y)-z\partial_\alpha w(x,y)$, where $\alpha$ stands for the in-plane directions $x$, $y$ and $v_\alpha$ describes the in-plane components of the displacement. In the regime considered in this work, the in-plane components of the displacement  $v_\alpha$ are negligible with respect to the out-of-plane components, as confirmed by FEM simulations, thus we neglect them \cite{Atalaya_2008}.\\
The stress and strain tensors in terms of the out-of-plane displacement are
\begin{align}
    \varepsilon_{\alpha\beta}(x,y)&=\varepsilon_0\delta_{\alpha\beta}-z\partial_{\alpha\beta}w+\frac{1}{2}\partial_\alpha w\partial_\beta w,\label{strain}\\
    \sigma_{\alpha\beta}(x,y)&=\frac{E}{1-\nu^2}\left((1-\nu)\varepsilon_{\alpha\beta}+\nu\varepsilon_{\gamma\gamma}\delta_{\alpha\beta}\right),
\end{align}
where the Greek indices stands for the in-plane directions $x$ and $y$, and we include a static in-plane deformation $\varepsilon_0(x,y)$, giving rise to a static stress $\sigma_0(x,y)$. This can be generalized in the time domain by assuming a time-dependent out-of-plane displacement $w(x,y,t)$.\\
A membrane deformed under the conditions described above behaves accordingly to the Von K\'{a}rm\'{a}n theory \cite{Landau1970}. In particular, we are interested in the dynamics that is described by the set of equations
\begin{equation}\label{eq:motion}
    \begin{cases}
    \rho h\ddot{w}-\partial_{\alpha\beta}M_{\alpha\beta}-\partial_{\beta}(N_{\alpha\beta}\partial_\alpha w)&=0\\
    \partial_\beta N_{\alpha\beta}=0
\end{cases}
\end{equation}
here we introduce the two stress resultants $N_{\alpha\beta}$ and $M_{\alpha\beta}$ which are defined as
\begin{align}
    N_{\alpha\beta}&=\int_{-\frac{h}{2}}^{\frac{h}{2}}\sigma_{\alpha\beta}dz\label{eq:shearforce}\\
    M_{\alpha\beta}&=\int_{-\frac{h}{2}}^{\frac{h}{2}} z\sigma_{\alpha\beta}dz\label{eq:bendingmoment}
\end{align}
Written in this way, the equation of motion doesn't present any type of dissipation. To include the dissipative terms we assume that there is a small, constant delay time between stress and strain
\begin{align*}\label{eq:stressdiss}
    \sigma_{ij}&=\frac{E}{1+\nu}\left(\varepsilon_{ij}(t-\tau)+\frac{\nu}{1-2\nu}\varepsilon_{kk}(t-\tau)\delta_{ij}\right)\\
    &\simeq \frac{E}{1+\nu}\left(\varepsilon_{ij}(t)+\frac{\nu}{1-2\nu}\varepsilon_{kk}(t)\delta_{ij}\right)+\\
    &-\frac{E\tau}{1+\nu}\left(\frac{\partial\varepsilon_{ij}(t)}{\partial t}+\frac{\nu}{1-2v}\frac{\partial\varepsilon_{kk}(t)}{\partial t}\delta_{ij}\right)\\
    &=\sigma_{ij}^\mathrm{cons}+\sigma_{ij}^\mathrm{diss}.
\end{align*}
Substituting eq. \eqref{eq:stressdiss} in eq. \eqref{eq:shearforce} and eq. \eqref{eq:bendingmoment}, we can write the stress resultants including the dissipative terms:
\begin{align}
    N_{\alpha\beta}^{\mathrm{cons}}&=\frac{Eh}{1-\nu^2}\left[(1-\nu)\delta_{\alpha\beta}\varepsilon_0+\frac{1-\nu}{2}\partial_\alpha w\partial_\beta w+\nu\partial_\gamma w\partial_\gamma w\right]\\
    N_{\alpha\beta}^{\mathrm{diss}}&=\frac{Eh}{1-\nu^2}\tau\left[\frac{1-\nu}{2}(\partial_\alpha \dot{w}\partial_\beta w+\partial_\alpha w\partial_\beta\dot{w})+\nu\partial_\gamma \dot{w}\partial_\gamma w\right]\\
    M_{\alpha\beta}^\mathrm{cons}&=-D\left[(1-\nu)\partial_{\alpha\beta} w+\nu\partial_{\gamma\gamma}w\delta_{\alpha\beta}\right]\\
    M_{\alpha\beta}^\mathrm{diss}&=-D\tau\left[(1-\nu)\partial_{\alpha\beta}\dot{w}+\nu\partial_{\gamma\gamma}\dot{w}\delta_{\alpha\beta}\right]
\end{align}
where we introduce the flexural rigidity $D=Eh^2/12(1-\nu^2)$.
Writing explicitely the stress resultant in terms of $w$, the solution of the system formed by eq. \eqref{eq:motion} is
\begin{equation}
\begin{split}
    &\rho h \ddot{w}+D(\partial_{\alpha\alpha\beta\beta }w+\tau\partial_{\alpha\alpha\beta\beta}\dot{w})-\frac{Eh\varepsilon_0}{1+\nu}\partial_{\alpha\alpha}w-\frac{Eh}{1-\nu^2}\left[\frac{1-\nu}{2}\partial_{\alpha\beta}w\partial_{\alpha}w\partial_\beta w+\partial_{\alpha\alpha}w\partial_\beta w\partial_\beta w\right]+\\
    &-\frac{Eh\tau}{1-\nu^2}\left[\frac{1-\nu}{2}(\partial_{\alpha\beta}w\partial_\alpha\dot{w}\partial_\beta w+\nu\partial_{\alpha\alpha} w\partial_\beta w\partial_\beta\dot{w}\right]=0.
    \end{split}
\end{equation}
Such equation describes the dynamic of the total displacement field $w$ and it presents some nonlinear terms. In the linear case, we can expand the total displacement field over a set of normalized eigenmodes $\phi_n$ associated with an out-of-plane oscillation $u_n$. 
For small nonlinearities, as the one considered in this work, we can assume that the set of eigenmodes obtained in the linear case form a basis over which we can expand the displacement field $w(x,y,t)=\phi_n(x,y)u_n(t)$. By performing this expansion the equation of motion becomes
\begin{equation}
\begin{split}
    &\rho h\phi_n\ddot{u}_n+\tau\left(D\partial_{\alpha\alpha\beta\beta}\phi_n\right)\dot{u}_n-\frac{Eh\tau}{1-\nu^2}\left[(1-\nu)\partial_{\alpha\beta}\phi_n\partial_\alpha\phi_n\partial_\beta\phi_n+\nu\partial_{\alpha\alpha}\phi_n\partial_{\beta}\phi_n\partial_\beta\phi_n\right]u_n^2\dot{u}_n+\\
    &+\left[D\partial_{\alpha\alpha\beta\beta}\phi_n-Eh\sigma_0\partial_{\alpha\alpha}\phi_n\right]u_n-\frac{Eh}{2(1-\nu^2)}\left[(1-\nu)\partial_{\alpha\beta}\phi_n\partial_\alpha\phi_n\partial_\beta\phi_n+\nu\partial_{\alpha\alpha}\phi_n\partial_{\beta}\phi_n\partial_\beta\phi_n\right]u_n^3=0
\end{split}    
\end{equation}
were we group together all the terms of the same order in $u_n$ and $\dot{u}_n$. Since we are interested in the dynamic of a single mode we apply a discretization method which allow us to describe the dynamic of a single mode in terms effective parameters. We apply the so-called Galerkin method, which consists in multiplying by a test function, i.e. the mode of interest $\phi_i$, and perform the integral over the membrane area
\begin{equation}
    \begin{split}
    &\ddot{u}_n\rho h\int_S\phi_i\phi_n dA+\dot{u}_n\tau\int_S \phi_i\left(D\partial_{\alpha\alpha\beta\beta}\phi_n\right)dA+\\
    &-u_n^2\dot{u}_n\frac{Eh\tau}{1-\nu^2}\int_S\phi_i\left[(1-\nu)\partial_{\alpha\beta}\phi_n\partial_\alpha\phi_n\partial_\beta\phi_n+\nu\partial_{\alpha\alpha}\phi_n\partial_{\beta}\phi_n\partial_\beta\phi_n\right]dA+\\
    &+u_n\int_S\phi_i\left[D\partial_{\alpha\alpha\beta\beta}\phi_n-Eh\sigma_0\partial_{\alpha\alpha}\phi_n\right]dA +\\
   & -u_n^3\frac{Eh}{2(1-\nu^2)}\int_S\phi_i\left[(1-\nu)\partial_{\alpha\beta}\phi_n\partial_\alpha\phi_n\partial_\beta\phi_n+\nu\partial_{\alpha\alpha}\phi_n\partial_{\beta}\phi_n\partial_\beta\phi_n\right]dA=0
    \end{split}
\end{equation}
We can now exploit the orthogonality property of the basis' vectors, i.e. $\int\phi_i\phi_ndA=0$ if $n\not= i$, and we perform a single mode approximation, i.e. we assume that all the combination of partial derivative of $\phi_i$ and $\phi_n$ are negligible if $n\not= i$. Within this assumption we can write the equation of motion as an effective equation
\begin{equation}
    \ddot{u}_i+\Gamma_i\dot{u}_i+\gamma_i^{\mathrm{nl}}u_i^2\dot{u}_i+\Omega_i^2u_i+\beta_iu_i^3=0
\end{equation}
where 
\begin{align}
    m_\mathrm{eff}&=\rho h \int_S\phi_i^2dA\\
    \Omega_i^2&=m_\mathrm{eff}^{-1}\int_S\phi_i\left(D\partial_{\alpha\alpha\beta\beta}\phi_i-Eh\sigma_o\partial_{\alpha\alpha}\phi_i\right)dA\\
    \Gamma_i&=m_\mathrm{eff}^{-1}D\tau\int_S\phi_i\partial_{\alpha\alpha\beta\beta}\phi_idA\\
    \beta_i&=-\frac{Eh}{2m_\mathrm{eff}(1-\nu^2)}\int_S\phi_i[(1-\nu)\partial_{\alpha\beta}\phi_i\partial_\alpha\phi_i\partial_\beta\phi_i+\nu\partial_{\alpha\alpha}\phi_i\partial_\beta\phi_i\partial_\beta\phi_i]dA\\
    \gamma_i^\mathrm{nl}&=-\frac{Eh\tau}{m_\mathrm{eff}(1-\nu^2)}\int_S\phi_i[(1-\nu)\partial_{\alpha\beta}\phi_i\partial_\alpha\phi_i\partial_\beta\phi_i+\nu\partial_{\alpha\alpha}\phi_i\partial_\beta\phi_i\partial_\beta\phi_i]dA
\end{align}

\section{Methods}\label{app:method}
We measure the displacement of the membrane using a fiber-based interferometer with an heterodyne receiver. We refer to the beam reflected by the mebrane as the {\it signal} beam, with optical power $P_s$. The reference beam, called {\it local oscillator}, has a power of $P_{LO}$, usually much stronger than the signal one, and is frequency-shifted by $40$~MHz.
The resulting photocurrent contains a dominant beat note at that frequency, and sidebands around it due to the phase modulation imparted on the signal beam by the mechanical motion. In addition, the photodetector has a response given by $D$, which we assume to be constant around the frequencies of interest.
We analyze this photocurrent by means of a lock-in amplifier (HF2LI Z{\"u}rich Instrument). In particular, we extract by demodulation both the frequency components around the carrier and mechanical sidebands of the mode of interest. We can combine the in-phase and quadrature components together to form the complex outcomes $z_\text{car}$ and $z_\text{sb}$, which satisfy
\begin{subequations}
  \label{eq:dem_outcomes}
\begin{align}
  \label{eq:5}
  z_\text{car}(t) &= H(t) * D \sqrt{P_\text{LO} P_s} J_0\left( G A_i(t)\right)\,e^{-\imath \phi(t)}, \\
  z_\text{sb}(t) &= \mp H(t) * D \sqrt{P_\text{LO} P_s} J_1\left( G A_i(t)\right)\,e^{\mp\imath \psi_i(t)}e^{\imath\phi(t)},
\end{align}
\end{subequations}
where $*$ is the convolution operator, $H(t)$ the kernel of the low-pass filter used in the demodulation process, $J_n$ the $n$th Bessel function of the first kind, $G:=4\pi \Lambda/\lambda$ the optomechanical coupling with $\lambda$ the laser wavelength and $\Lambda$ the transverse optomechanical modal overlap integral, $\phi$ is the relative phase fluctuations between the local oscillator and signal beams. In addition, $A_i(t)$ and $\psi_i(t):=\int^t_0 \delta \Omega_i(s)ds$ are, respectively, the displacement amplitude and instantaneous phase of the mechanical mode of interest, as defined in eqs.~(10) and (11) of the main text.

\subsection{Displacement calibration}
The demodulated outcomes $z$ are usually in electrical units, $V_\text{rms}$. In many cases it is desirable to calibrate the results in absolute displacement units.
In principle, this calibration can be accomplished by independently characterising each term in eq.~\eqref{eq:dem_outcomes}.
In practice however, there are always uncertainties in some of the parameters, especially the modal overlap integral $\Lambda$.
Then, we employ a different calibration technique based on thermometry of mechanical motion.
We record time traces for both the carrier and the upper sideband at the thermal equilibrium, i.e. without any external force exerted to the membrane. In this case, the mechanical modulation depth induced by the thermal motion is small, that is $GA_i\ll 1$, then eqs.~\eqref{eq:dem_outcomes} can be approximated to first order as 
\begin{subequations}
  \label{eq:dem_outcomes_lin}
\begin{align}
  \label{eq:5}
  z_\text{car}(t) &\approx H(t) * D \sqrt{P_\text{LO} P_s} \,e^{-\imath \phi(t)}, \\
  z_\text{sb}(t) &\approx \mp H(t) * D \sqrt{P_\text{LO} P_s}  G A_i(t)/2\,e^{\mp\imath \psi_i(t)}e^{\imath\phi(t)}.
\end{align}
\end{subequations}
According to the previous equations, we use the amplitude of the carrier to normalize the sideband outcome, that is $y=2z_\text{sb}/|Z_\text{car}|$. In this way, we remove the demodulation filter, the detector gain and any drifts in the optical power.
In order to obtain the optomechanical coupling $G$, we calculate the power spectral density of the measurement record $y$, then we integrate it around the Lorentzian of the mechanical mode in order to obtain an estimate of the mechanical displacement $\Delta y^2 = G^2\langle A_{i,th}^2\rangle$. Finally, the equipartition theorem assures that $\langle A_{i,th}^2\rangle = k_B T/(m_\text{eff} \Omega_i^2)$, where $k_B$ is the Boltzmann's constant and $T$ the membrane mode temperature, which we assume to be equal to the lab temperature of 294 K. Assuming the effective mass from simulations \cite{Tsaturyan_2017}, we can obtain the optomechanical coupling as $G=\sqrt{\Delta y^2 / A_{i,th}^2}$.

\subsection{Nonlinear transduction}
Within the linear regime of eqs.~\eqref{eq:dem_outcomes_lin}, the demodulated sideband is proportional to the amplitude $A_i$, thus can be directly use as a displacement measurement.
However for nonlinear ringdown measurements, the amplitude is excited to a large value such that the linear approximation made above is no longer valid, and the nonlinear transduction has to be properly take into account.
To do that, we notice that in the nonlinear regime the mechanical modulation depth becomes comparable to unity, i.e. $G A_i\sim1$ and the carrier's amplitude starts to reduce. At the beginning of the ringdown this modulation depth is maximum, and the nonlinear transduction effects dominate. As the mechanical displacement decays, the associate modulation depth reduces and the transduction recovers the usual linear regime.
In order to undo this nonlinear effect, we first normalize the carrier's magnitude $|z_\text{car}|$ to the value obtained in the linear regime. This directly provides us with a measurement of $\overline{z}=J_0\left(GA_i(t)\right)$, which we fit using the function $1 - b e^{-t/\tau}$. An example is shown in Fig.~\ref{f:SI_nonlin-trans}a.
\begin{figure}[ht!]
\center
\includegraphics[scale=.9]{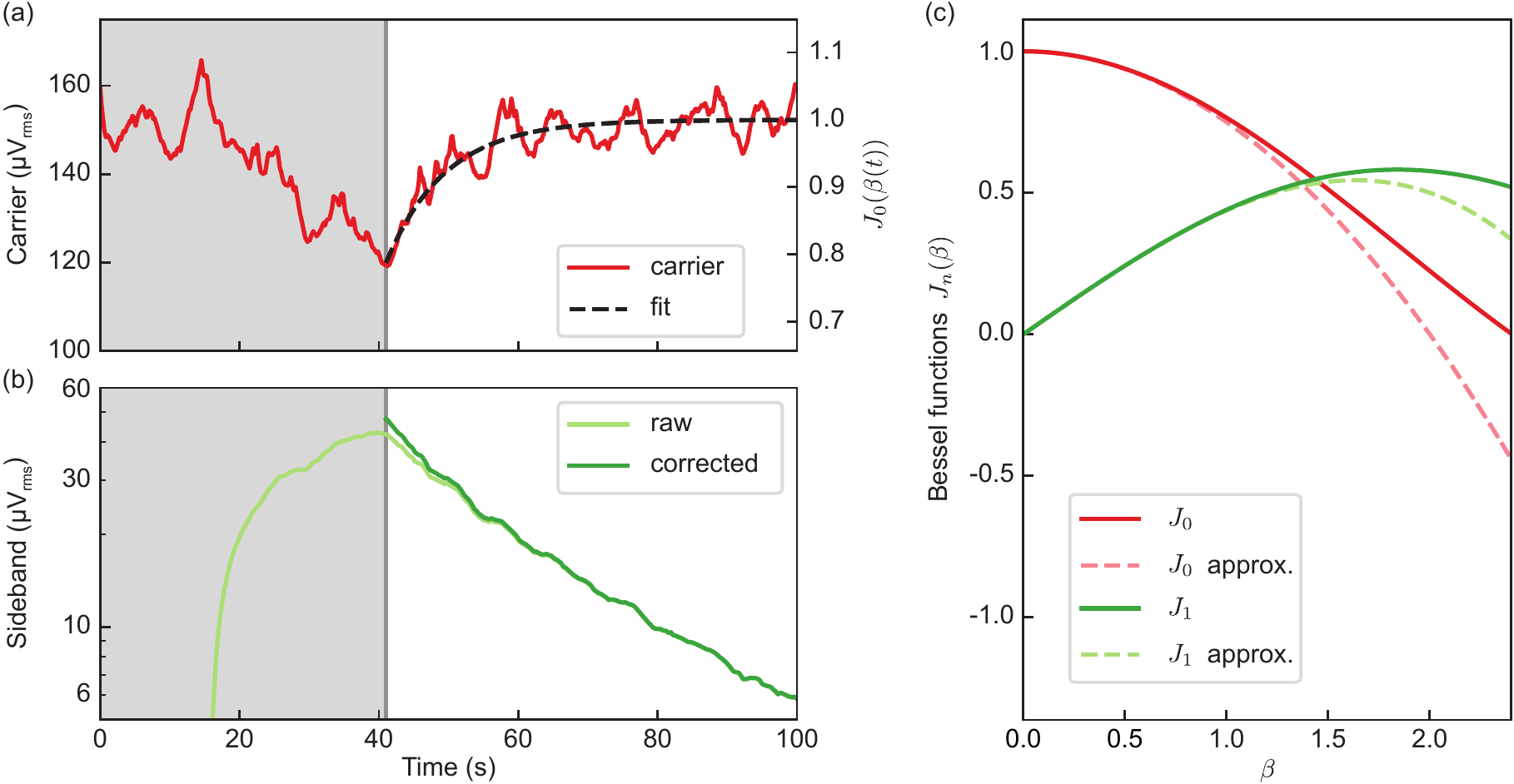}
\caption{Accounting for interferometric nonlinear transduction.
  (a) Demodulated carrier signal (red), during excitation and subsequent ringdown. The ring-up during the decay is fitted with an exponential function (black) and is due to transduction nonlinearity. Upon normalization to the steady state, the value of the Bessel function is retrieved (right axis).
  (b) Demodulated sideband signal (light green), during excitation and subsequent ringdown. The dark green trace corresponds to the corrected measurement, where the effect of nonlinear transduction is undone. Excess nonlinear damping is revealed at the beginning of the decay.
  (c) Zeroth (solid red) and first (solid green) Bessel function, as well as approximation to the third order (dashed lines).
}
\label{f:SI_nonlin-trans}
\end{figure}
Next, we invert the Bessel function to obtain the modulation depth $\beta=J_0^{-1}(\overline{z})$ and use it to correct the nonlinear transduction induced in the sideband outcome, that is
\begin{eqnarray}
  \label{eq:nonlin-corr-1}
  z_\text{sb}^\text{corr} = z_\text{sb} \frac{\beta/2}{J_1(\beta)},
\end{eqnarray}
where the factor $\beta/2$ corresponds to the linear approximation of the first Bessel function. An example of ringdown with this correction is shown in Fig.~\ref{f:SI_nonlin-trans}.
We notice that in the displacement regime under which the experiment is operated, the first two Bessel functions are always positive and invertible.

In our experiments, the achieved modulation depths satisfy $\beta \leq 1$. In this case, the Bessel functions are well approximated by an expansion to the third order (see Fig.~\ref{f:SI_nonlin-trans}c), that is
\begin{subequations}
  \label{eq:app-bessel-3order}
  \begin{align}
    J_0(\beta)&= 1 - \left(\frac{\beta}{2}\right)^2 + o(\beta^4),\\
    J_1(\beta)&= \frac{\beta}{2}-\frac{1}{2}\left(\frac{\beta}{2}\right)^3 + o(\beta^5)=\frac{\beta}{2}\sqrt{J_0(\beta)} + o(\beta^4),\label{eq:app-bessel-1}
  \end{align}
\end{subequations}
where in the last equality we have used $\sqrt{J_0(\beta)}\approx\sqrt{1-(\beta/2)^2}\approx1-(\beta/2)^2/2$. In practice, eq.~\eqref{eq:app-bessel-1} suggests that an alternative way to undo nonlinear transduction effects is to divide out the measured normalized carrier, $\overline{z}$, from the sideband, that is
\begin{eqnarray}
  \label{eq:nonlin-corr-2}
  z_\text{sb}^\text{corr} = z_\text{sb} \frac{1}{\sqrt{\overline{z}}}.
\end{eqnarray}
Indeed, we find excellent agreement between the methods in eq.~\eqref{eq:nonlin-corr-1} and eq.~\eqref{eq:nonlin-corr-2}.

\subsection{Instantaneous frequency shift}
In order to extract the Duffing parameter, we need to estimate the instantaneous frequency of the mechanical mode, during the ringdown. To do that, we first obtain the phase $\psi_i+\phi$ from the measured sideband $z_\text{sb}$. This phase contains both the mechanical one, $\psi_i$, and the relative phase  between the two arms of the interferometer, $\phi$. This relative phase, in general, is subjected to fluctuations and slow drifts, as the interferometer is not actively locked. Nevertheless, this phase is recorded in the carrier outcome, $z_\text{car}$, from which we get $\phi$ and remove it to the phase of the sideband to finally get the mechanical phase $\psi_i$.
Next, we perform a numerical derivative in order to estimate the frequency shift $\delta\Omega_i$. The obtained frequency shift, shown for instance in Fig.~2d of the main text, is fitted to a polynomial function $a + b A_i^2 + c A_i^4$, where $b$ is the Duffing term and we have introduced a quartic correction.

\section{Complete set of data}\label{app:data}
Here we report the complete set of nonlinear loss angle measured. In order to collect statistic, we perform the same measurement on several membrane nominally identical. On each membrane we repeat the measurement on four bandgap modes. For each membrane we perform repeated measurement on a single mode to better estimate the parameters. All the fits giving a parameter with an error larger than 10$\%$ have been discarded. The error is extracted form the $95\%$ confidence interval. For the mode 3 in two membranes we observed negative Duffing terms. The complete set of data is shown in Fig.~\ref{f:figS1}. 
\begin{figure*}[ht!]
\center
\includegraphics[scale=.9]{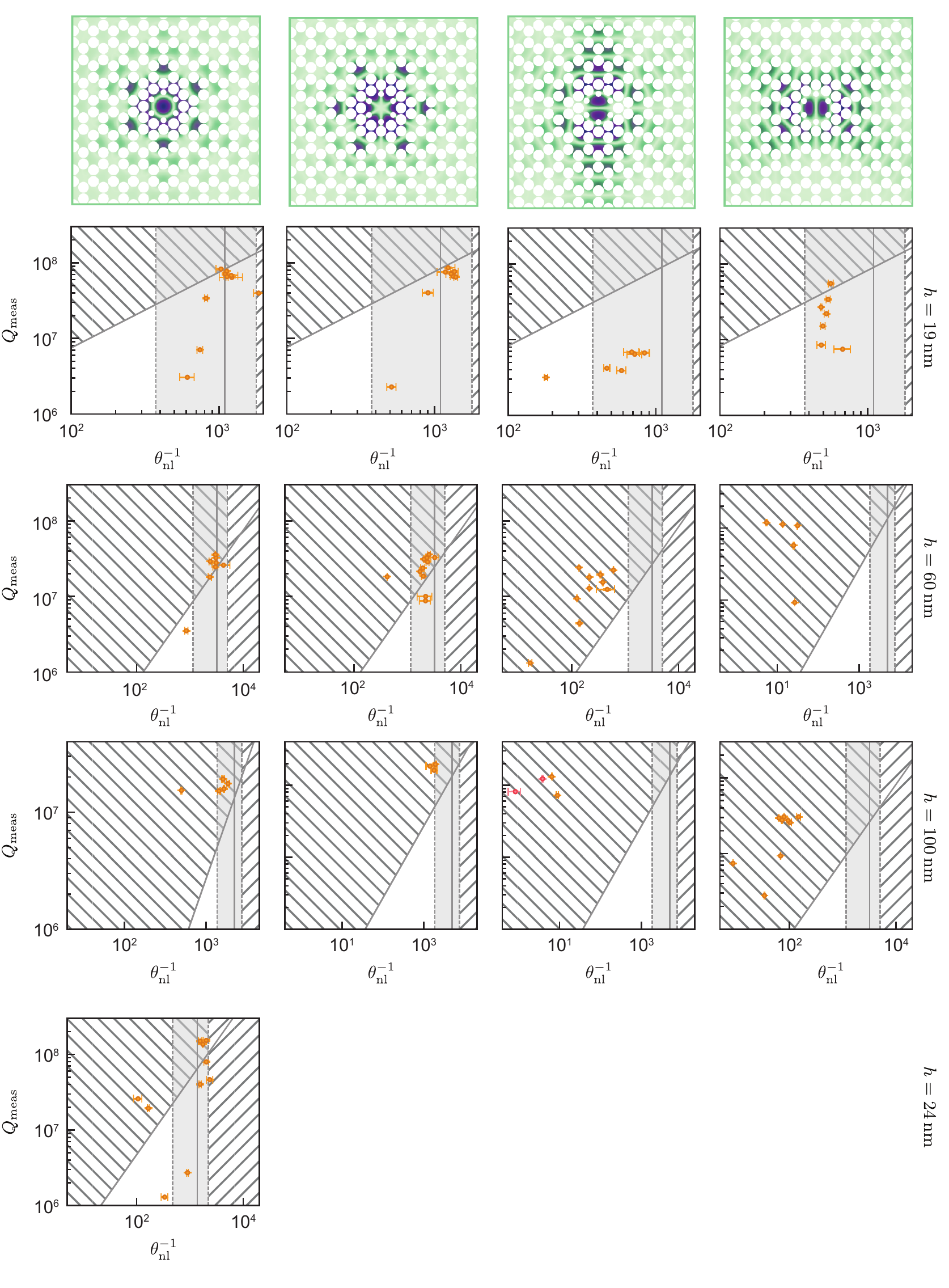}
\caption{Complete set of measured nonlinear loss angles. Each panel shows all nonlinear loss angles measured for one mode (showed on top) on several membranes with the same thickness (showed on the right). Each point corresponds to the mean value over 5 ringdowns performed on the same membrane. The two red points (mode 3, $h=100$ nm) correspond to the membranes where we observed a negative duffing shift. For those we plot $-\theta_\mathrm{nl}^{-1}$.
}
\label{f:figS1}
\end{figure*}
\end{widetext}
\clearpage

%

\end{document}